\documentclass[pre,showpacs,amssymb,superscriptaddress,floatfix,a4paper,twocolumn]{revtex4}
\usepackage{graphicx}
\usepackage{bm}
\usepackage{amsmath}

\begin{document}

\title{Effective Edwards-Wilkinson equation for single-file diffusion}

\author{P. M. Centres}
\affiliation{Departamento de F\'{\i}sica, Instituto de F\'{\i}sica Aplicada, Universidad Nacional de San Luis-CONICET, Chacabuco 917, D5700HHW, San  Luis, Argentina}

\author{S. Bustingorry}
\affiliation{CONICET, Centro At{\'{o}}mico Bariloche, 8400 San Carlos de Bariloche, R\'{\i}o Negro, Argentina}

\date{\today}

\begin{abstract}

In this work, we present an effective discrete Edwards-Wilkinson equation aimed to describe the single-file diffusion process. The key physical properties of the system are captured defining an effective elasticity, which is proportional to the single particle diffusion coefficient and to the inverse squared mean separation between particles. The effective equation gives a description of single-file diffusion using the global roughness of the system of particles, which presents three characteristic regimes, namely normal diffusion, subdiffusion and saturation, separated by two crossover times. We show how these regimes scale with the parameters of the original system. Additional repulsive interaction terms are also considered and we analyze how the crossover times depend on the intensity of the additional terms. Finally, we show that the roughness distribution can be well characterized by the Edwards-Wilkinson universal form for the different single-file diffusion processes studied here.

\end{abstract}


\maketitle

\section{Introduction}
\label{sec:intro}

Single file diffusion (SFD) is one of the simplest and non-trivial problems in statistical physics~\cite{harris1965,fedders1978,alexander1978,beijeren1983}. In this problem a one-dimensional collection of particles diffuses with hard-core repulsion as the basic constraint. In one dimension, hard-core repulsion implies the non-passing rule, i.e. a given particle can not cross over its neighbor particles. This restricts the phase space generating an inherent entropic interaction. Despite its simplicity the SFD problem still attracts much attention due mostly to the possibility of applying different analytical formalisms~\cite{lin2005,lizana2008,lizana2009,taloni2008,barkai2009}.

A direct experimental observation of SFD processes has become possible during the last decade. The improvement of experimental techniques has permitted studying systems ranging from interacting colloidal particles on channels or rings at the microscopic scale~\cite{wei1999,wei2000}, to charged metallic balls on rings at the macroscopic scale~\cite{coupier2006}. Other interesting problems in which single-file diffusion is important are the longitudinal relaxation of a single polymer~\cite{amitai2010,doi1986} or the transverse fluctuation of vecinal surfaces~\cite{giesen2001}.

The simple hard-core interaction gives rise to non-trivial diffusion properties. If the diffusion of a tagged particle is followed, it first presents a normal diffusion regime, i.e. $\Delta(t)=\langle [r(t) - \langle r \rangle(t)]^2 \rangle = 2Dt$, where $r(t)$ is the instantaneous position of the particle, $\langle \dots \rangle$ is an ensemble average, and $D$ is the diffusion coefficient. This relation is valid up to a characteristic time, beyond which, in average, particles start colliding with their neighbors. Then, as a consequence of the interaction between particles and the one-dimensional character of the problem, the tagged particle shows a subdiffusive behavior of the form $\Delta(t) = 2Ft^{1/2}$, where $F$ is a characteristic coefficient. This subdiffusive regime last for ever in infinite systems, but it crosses over to saturation or normal diffusion in finite systems, depending on the applied boundary conditions~\cite{beijeren1983}. The subdiffusive regime of a single tagged particle is the signature of the interacting system. On the contrary, the diffusion of the center of mass is always normal, i.e. it increases linearly with time, without showing any signature of the interacting system.

Different analytical approaches to the SFD problem have been recently proposed. Among others, we can mention a Langevin formulation~\cite{taloni2008}, the calculation of the probability density function for the position of a tagged particle using a Bethe ansatz~\cite{lizana2008,lizana2009}, an approach based on fractional Brownian motion with a variable Hurst exponent~\cite{lim2009}, or a formulation for the tagged particle motion using classical reflection and transmission coefficients~\cite{barkai2009}. In the present work, we propose a different and complementary approach based on an effective description of the SFD problem in terms of a discrete Edwards-Wilkinson equation (DEWE). The two parameters of the DEWE are the noise intensity and the elasticity of the system. In order to present an effective equation, effective values for these parameters are given, relating them to the SFD parameters, i.e. to the mean particle separation and to the single particle diffusion coefficient. Then, we describe the SFD process using the global roughness of the system of particles, which gives information on how the particles diffusion becomes correlated with time. It is shown that the description using the DEWE successfully accounts for the case of a pure hard-core interaction. In order to extend our analysis, we consider two further cases of interacting particles, one with a purely harmonic repulsion term, and the other with an interaction term proportional to the reciprocal of the squared distance between two consecutive particles. Finally we discuss some aspects of the universal properties of the distribution function of the global roughness.

\section{Model system}
\label{sec:model}

\begin{figure}[!tbp]
\includegraphics[width=8.5cm,clip=true]{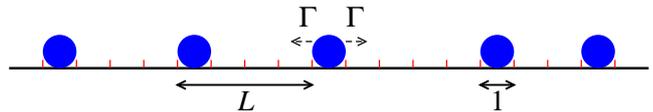}
\caption{\label{f:model}(Color online) Schematic representation of the model single-file diffusion system. Each particle of unitary size diffuses in a discrete one-dimensional system with periodic boundary conditions. The mean separation between particles is $L$. Each particle can move with equal transition probability $\Gamma =1/2$ to one of its next nearest neighbor sites if they are not occupied by other particle.}
\end{figure}

Let us first briefly describe the model system under study. We consider a set of $N$ particles of unitary size in a discrete one-dimensional system. The particles are in average a distance $L$ apart and thus the size of the system is $LN$ and the particle density is $\rho = 1/L$, as depicted in Fig.~\ref{f:model}. For the sake of simplicity, we use periodic boundary conditions. The $j$ label is used for the particles, so the instantaneous position for the $j\text{-th}$ particle at time $t$ is $r_j(t)$. Instead of this instantaneous particle position, let us consider the displacement field $u_j(t)=r_j(t)-jL$, which measures the displacement with respect to an ideal ordered lattice $r_j = j L$.

We shall consider the motion of the particles in a discrete medium. The dynamics of each particle consists of a random displacement to one of its two neighboring sites, with equal transition probabilities $\Gamma_{u,u+1}=\Gamma_{u,u-1}=\Gamma=1/2$, only if the new site is not occupied by one of the neighboring particles. This means that the hard-core repulsion restricts the position of the $j$ particle to $r_{j-1}(t) < r_j(t) < r_{j+1}(t)$ for all times. This represents indeed an entropic repulsion between particles. Note that, expressed in terms of the displacement field, the non-passing rule may be written as
\begin{equation}
\label{eq:non-pass}
 u_{j-1}(t) - L < u_j(t) < u_{j+1}(t) + L.
\end{equation}

Within this simple model, the interacting particles follow a SFD process. Thence the single particle diffusion shows three characteristic regimes: a single particle normal diffusion regime $\Delta = \langle [r_j(t)-\langle r_j(t) \rangle ]^2 \rangle \sim t$, an interacting particles subdiffusive regime $\Delta \sim t^{1/2}$, and a center-of-mass normal diffusion regime $\Delta \sim t$. The subdiffusive regime is the key signature of the interaction between particles.

As described here, this simple model with hard-core repulsion and constant diffusion coefficient $\Gamma$ does not depend on temperature. One can think of other interaction terms which allow for a temperature dependence. For instance, by including a simple harmonic repulsion between particles given by the Hamiltonian
\begin{equation}
 H_K = \frac{K}{2} \sum_{j=0}^{N-1} \left[ u_{j+1}(t) - u_j(t) + L \right]^2.
\end{equation}
Periodic boundary conditions are implicitly assumed in the sum. We shall analyze the model containing the hard-core repulsion plus the harmonic contribution as the simplest model with an energetic (non-entropic) interaction term. We shall also consider another energetic interaction term where the interaction is proportional to $1/r^2$, being $r$ the distance between two consecutive particles. We are interested in this interaction form since it is relevant for the vecinal surface problem~\cite{giesen2001}. In such case, the Hamiltonian reads
\begin{equation}
 H_A = A \sum_{j=0}^{N-1} \left[ u_{j+1}(t) - u_j(t) + L \right]^{-2}.
\end{equation}

In order to compare with the effective DEWE description of the process we present in the following results from direct numerical simulations of the present model. A Monte Carlo scheme can be implemented to perform numerical simulations of the model described so far. For the single-file diffusion mechanism a random particle is selected and one of its two neighbors randomly chosen with equal probability, and then the particle moves only if the new site is empty, which implements the non-passing rule Eq.~\eqref{eq:non-pass}. When an energetic contribution is considered, it is taken into account in the dynamics of the particles with a Metropolis rule. The new site for a given particle is randomly selected as before, then the energy difference $\Delta H$ is considered and the probability $P = \exp(-\Delta H/T)$ is computed. The new particle moves to the new site if $P > \xi$, where $\xi$ is a random number uniformly distributed in $(0,1]$. This Monte Carlo scheme will be used to compare with the effective description of the model in terms of the DEWE, which is presented in the following section.

\section{Discrete Edwards-Wilkinson equation}

In this section we present the effective DEWE which allows for a description of the main physical properties of the SFD problem. We shall first present the well-known general solution of the equation and we will then discuss how its parameters can be effectively related to the parameters of the interacting particle system. Notice that although we use here a continuous-time Edwards-Wilkinson equation for the sake of simplicity, the results we will present here can be directly translated to the discrete-time version.

\subsection{General results}

\begin{figure}[!tbp]
\includegraphics[width=8.5cm,clip=true]{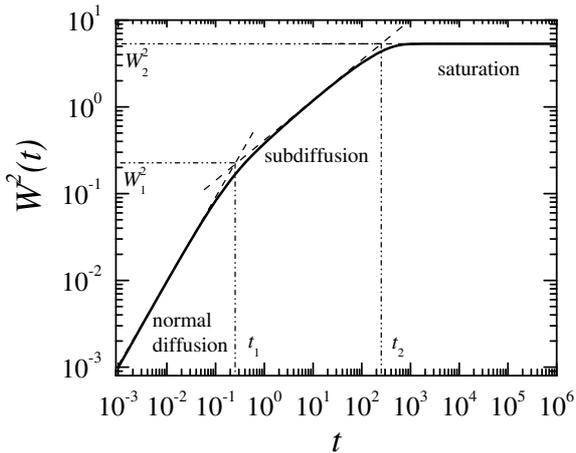}
\caption{\label{f:w2-regimes}Roughness of the DEWE for parameters $N=128$, $\nu=1$ and $D=1/2$. Dashed lines correspond to the three asymptotic regimes, which are separated by the crossover points $y_1=(t_1,W^2_1)$ and $y_2=(t_2,W^2_2)$, as shown in the figure.}
\end{figure}

The Edwards-Wilkinson equation~\cite{edwards1982} is characterized by the amplitude of the thermal noise $D$ and the elastic constant $\nu$. In a discrete form it can be written as
\begin{equation}
\label{e:dew}
 \partial_t u_j(t) = \nu \left[ u_{j+1}(t) + u_{j-1}(t) -2\,u_j(t) \right] + \eta_j(t),
\end{equation}
with $j=0,...,N-1$ and periodic boundary conditions. The noise term has a zero mean and the correlations $\langle \eta_j(t) \eta_{j'}(t') \rangle= 2 D \delta_{j,j'} \delta(t-t')$ have intensity $D$. The average over the white noise is denoted by angular brackets. Notice that we write the DEWE directly in terms of the displacement field variable, thus elastically connecting a system of $N$ interacting beads. We consider the center of mass of the displacement field as
\begin{equation}
 \overline{u(t)} = \frac{1}{N} \sum_{j=0}^{N-1} u_j(t)
\end{equation}
which serves to define the Fourier transform
\begin{equation}
\label{e:dewF}
 c_n(t) = \sum_{j=0}^{N-1} \left[ u_j(t) - \overline{u(t)} \right] e^{-i k_n j},
\end{equation}
with $k_n = 2 \pi n /N$. Then the DEWE is transformed to a set of independent equations for the time evolution of the Fourier modes,
\begin{equation}
 \partial_t c_n(t) = - \nu_n c_n(t) + \eta_n(t),
\end{equation}
with
\begin{equation}
\label{e:defnun}
 \nu_n = 4 \nu \sin^2 (\pi n/N).
\end{equation}
The sinusoidal character of this damping constant is associated to the discreteness of the original equation. When one considers the continuum Edwards-Wilkinson equation $\nu_n \to \nu (2\pi n/N)^2 = \nu k_n^2$. The solution to the set of equations for $c_n$ reads
\begin{equation}
 c_n(t) = c_n(0) e^{- \nu_n t} +\int_0^t dt' \, e^{-\nu_n (t-t')} \eta_n(t).
\end{equation}
From now on we shall consider the flat initial condition $c_n(0)=0$, the generalization to arbitrary initial conditions being straightforward.

The global roughness of the system is defined as
\begin{equation}
 W^2(t) = \left\langle \frac{1}{N} \sum_{j=0}^{N-1} \left[ u_j(t) - \overline{u(t)} \right]^2 \right\rangle,
\end{equation}
which in terms of the Fourier modes can be written as
\begin{equation}
 W^2(t) = \frac{1}{N} \sum_{j,n,m} \left\langle c_n(t) c_m(t) \right\rangle e^{i(k_n+k_m) j}.
\end{equation}
Finally, giving that
\begin{equation}
 \left\langle c_n(t) c_m(t) \right\rangle = \frac{2D}{N} \frac{\delta_{n,-m}}{\nu_n+\nu_m}\left[ 1-e^{-(\nu_n+\nu_m)t} \right],
\end{equation}
one arrives at the result for the global roughness of the system,
\begin{equation}
\label{e:w2DEW}
 W^2(t) = \frac{D}{N} \sum_{n=1}^{N-1} \frac{1-e^{-2 \nu_n t}}{\nu_n},
\end{equation}
which includes the time dependence, the finite size $N$, and the discreteness of the system through $\nu_n$ given by Eq.~\eqref{e:defnun}.

The complete solution of the roughness, Eq.~\eqref{e:w2DEW}, has three time regimes. In the context of surface growth, where the Edwards-Wilkinson equation was originally proposed~\cite{barabasi-stanley}, the three regimes correspond to random deposition of particles on a substrate, correlated growth of the surface, and roughness saturation due to boundary conditions. The different regimes are described by
\begin{equation}
\label{eq:w2nu}
 W^2(t) = \left\{
 \begin{array}{lll}
  2Dt&\text{for}&t \ll t_1 \\
  D \sqrt{\frac{2t}{\pi \nu}}&\text{for}& t_1 \ll t \ll t_2 \\
  \frac{DN}{12 \nu}&\text{for}&t_2 \ll t
 \end{array}
\right.
\end{equation}
with the crossover times given by
\begin{subequations}
\label{eq:ynu1}
\begin{equation}
 t_1 = \frac{1}{2 \pi \nu}, 
\end{equation}
\begin{equation}
 t_2 = \frac{\pi N^2}{288 \nu}.
\end{equation}
\end{subequations}
The roughness associated to this crossover times are consequently given by
\begin{subequations}
\label{eq:ynu2}
\begin{equation}
  W^2(t_1)= W^2_1 = \frac{D}{\pi \nu}, 
\end{equation}
\begin{equation}
 W^2(t_2)= W^2_2 = \frac{DN}{12 \nu}.
\end{equation}
\end{subequations}

Figure~\ref{f:w2-regimes} depicts the different discussed regimes for the roughness obtained with the DEWE, Eq.~\eqref{e:w2DEW}. The solid line represents the roughness of a system with size $N=128$, elastic constant $\nu=1$ and noise intensity $D=1/2$. As can be observed, the three characteristic regimes are present, with the asymptotic forms indicated with dashed lines. Also shown in the figure are the two crossover points $y_1=(t_1,W^2_1)$ and $y_2=(t_2,W^2_2)$.

\subsection{SFD effective elasticity}

\begin{figure*}[!tbp]
\includegraphics[width=16cm,clip=true]{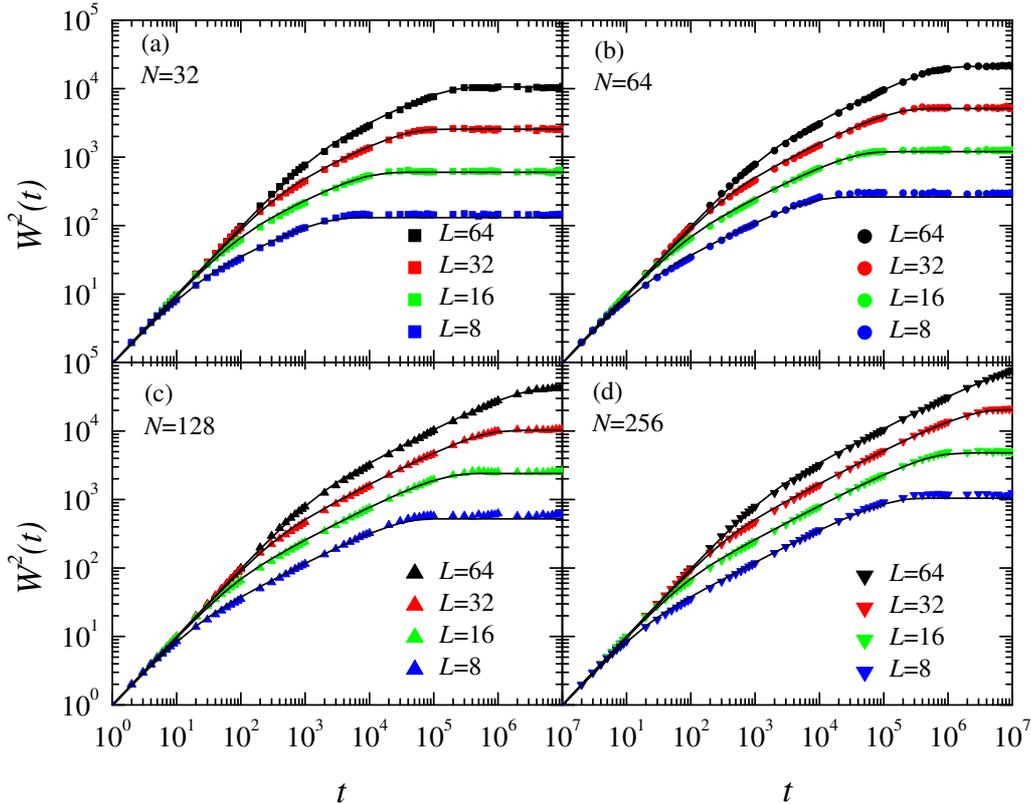}
\caption{\label{f:w2-num}(Color online) Roughness of the SFD problem. Points correspond to numerical simulation while continuous lines correspond to the DEWE. Different panels are for (a) $N=32$, (b) $N=64$, (c) $N=128$, and (d) $N=256$. In all panels, the different values of $L$ are indicated.}
\end{figure*}

In order to use the DEWE to describe the SFD problem, we have to relate its parameters with those of the SFD model presented in Sec.~\ref{sec:model}. Since the noise intensity should describe the diffusion of a single particle, it is directly mapped to the single particle diffusion coefficient, thus setting $D \equiv \Gamma$. Regarding the effective value for the elasticity $\nu$, it can in principle depend on both, the mean separation between particles $L$, and the diffusion coefficient $D$. One can argue that $\nu$ should increase with decreasing $L$, since the particles tend to see each other more frequently, thus further restricting the fluctuation of its neighbors. It can also be argued that $\nu$ should be proportional to the diffusion coefficient, i.e. for larger values of $D$ the system becomes more rigid. This is also akin to the compressibility of a simple gas, which is proportional to the temperature of the system. As we will show below in detail, the value
\begin{equation}
\label{eq:nuSFD}
 \nu_{\text{SFD}}=\frac{D}{(L-1)^2},
\end{equation}
gives a very good description of the SFD problem in terms of the DEWE. Note that in the last equation we have used the actual distance between two consecutive particles $L-1$ (see Fig.~\ref{f:model}), which takes into account the size of the particle, that in the present case is equal to one. This is important for small separations, but for the case $L \gg 1$, one has that $\nu_{\text{SFD}} \sim 1/L^2$.

One can now use the general results given in the last subsection taking $D=\Gamma$ and $\nu = \nu_{\text{SFD}}$. The next point to be considered is that the flat initial condition in the DEWE, $u_j(0)=0$, corresponds to a system of particles which are initially located in an ordered structure, i.e. $r_j(0)=jL$. For future reference in the present work, let us recall that using $\nu_{\text{SFD}}$, and from Eq.~\eqref{eq:w2nu}, the three regimes of the roughness for the SFD problem read now
\begin{equation}
 W^2(t) = \left\{
 \begin{array}{lll}
  2\Gamma t&\text{for}&t \ll t_1 \\
  (L-1)\sqrt{\frac{2 \Gamma t}{\pi}}&\text{for}& t_1 \ll t \ll t_2 \\
  \frac{N(L-1)^2}{12}&\text{for}&t_2 \ll t
 \end{array}
\right.
\end{equation}
with the crossover parameters given by
\begin{subequations}
\label{eq:y0}
\begin{equation}
t_1 = \frac{(L-1)^2}{2 \pi \Gamma} 
\end{equation}
\begin{equation}
 W^2_1 = \frac{(L-1)^2}{\pi} 
\end{equation}
\begin{equation}
 t_2 = \frac{\pi N^2 (L-1)^2}{288 \Gamma} 
\end{equation}
\begin{equation}
 W^2_2 = \frac{N (L-1)^2}{12}.
\end{equation}
\end{subequations}

It follows that for the SFD problem the three roughness regimes should be interpreted as follow. For $t \ll t_1$ single particles move independently, and thus this regime corresponds to normal diffusion. The first crossover time marks the time needed for a particle to diffuse over a distance of the order of the mean particle separation. Then, for times larger than $t_1$ particles start colliding with each other and the diffusion of the particles becomes correlated. Therefore, the second regime, $t_1 \ll t \ll t_2$, corresponds to the subdiffusive motion of particles. In this regime, there exists a growing correlation length which characterizes the number of particles whose motion is correlated at time $t$. The crossover time $t_2$ indicates when this correlation length reaches the size of the system and all particles ``know'' about the presence of all the other particles. Therefore, for $t \gg t_2$ the size of the system restricts the growth of the roughness of the system, thus corresponding to the saturation regime.

Figure~\ref{f:w2-num} shows numerical results for the dynamic global roughness of the SFD model together with the analytical roughness obtained with the DEWE using the effective elasticity $\nu_{\text{SFD}}$. Each panel corresponds to a different number of particles $N$, and for each $N$ different $L$ values are shown. In Fig.~\ref{f:w2-num}, symbols correspond to numerical simulations of the interacting particle system described in Sec.~\ref{sec:model}, while the continuous lines correspond to Eq.~\eqref{e:w2DEW} with $\nu=\nu_{\text{SFD}}$ and $D=\Gamma$. As can be observed, the proposed form for the elasticity does describe the numerical data accurately, specially for large values of $L$.

\begin{figure}[!tbp]
\includegraphics[width=8.5cm,clip=true]{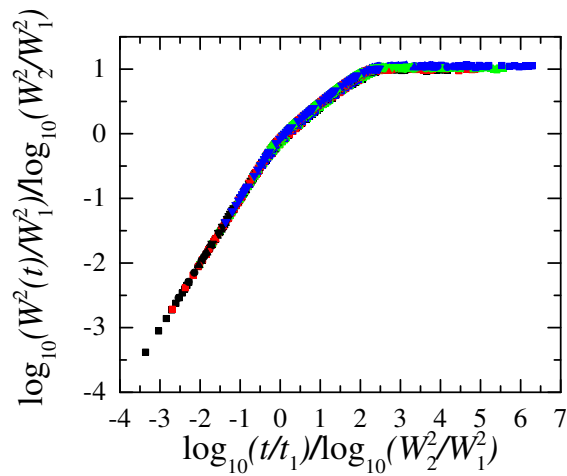}
\caption{\label{f:scal-chou}(Color online) Scaling of the dynamic global roughness shown in Fig.~\ref{f:w2-num} for different values of the SFD parameters.}
\end{figure}

\subsection{Roughness scaling function}

All the different curves shown in Fig.~\ref{f:w2-num}, also described through the DEWE, can be cast into a single universal form depicting the three regimes. When analyzing surface growth problems, the roughness is commonly scaled in terms of the Family-Vicsek form~\cite{family1985,barabasi-stanley}. However, this very usefull scaling form is well suited for roughness presenting two characteristic regimes, as the correlated growth and saturation regimes. Here, we are dealing with three regimes and thus the Family-Vicsek scaling form can not collapse all the data simultaneously. The problem of properly scaling a dynamic roughness with three regimes has been recently addressed by Chou and Pleimling~\cite{chou2009}. Based on a phenomenological construction, it has been shown that, once the two crossover points for each curve are estimated, the scaling can be described through
\begin{equation}
\label{eq:scal-chou}
 \frac{\log \left( \frac{W^2}{W^2_1} \right)}{\log \left( \frac{W^2_2}{W^2_1} \right)} =
 F\left[ \frac{\log \left( \frac{t}{t_1} \right)}{\log \left( \frac{W^2_2}{W^2_1} \right)} \right],
\end{equation}
where $F(x)$ is a given scaling function containing the three characteristic regimes. In order to use this scaling form the two crossover points must be estimated with all the prefactors. It is not enough to only know just its functional dependence on the key parameters of the problem.

Figure~\ref{f:scal-chou} shows how all the data contained in Fig.~\ref{f:w2-num} can be collapsed into a single universal function according to the proposed form, Eq.~\eqref{eq:scal-chou}. The data have been collapsed using the theoretical values for the crossover points, Eqs.~\eqref{eq:y0}, and not the numerical estimates of the crossover points. This stresses the relevance of the DEWE solution for the SFD problem.


\section{Additional Interaction terms}

In this section we analyze the effects of an additional interaction between particles. We shall consider two different energetic contributions giving rise to repulsive interactions, namely a purely harmonic term of intensity $K$ and a term including a repulsive interaction decaying with the inverse of the squared distance with intensity $A$. It is important to understand the effects of an harmonic term since, in principle, any interaction term can be reduced to the harmonic case in a strong interaction limit. On the other hand, the other interaction term is commonly used in the vecinal surface system~\cite{giesen2001} and will be analyzed as a typical non-trivial example.

\subsection{Roughness}

\begin{figure}[!tbp]
\includegraphics[width=8.5cm,clip=true]{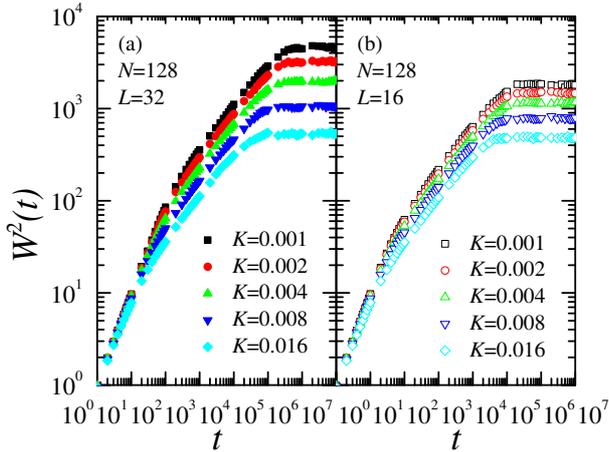}
\caption{\label{f:w2deK}(Color online) Behavior of the global roughness for the case of an additional harmonic interaction term and for $N=128$. The mean separation between particles is $L=32$ in panel (a) and $L=16$ in panel (b), and the different $K$ values are indicated.}
\end{figure}

Let us first show the general trend of the roughness with different interaction intensities. Figure~\ref{f:w2deK} shows the behavior of the roughness in the harmonic case and for a fixed system size $N=128$. The intensity of the harmonic term varies in the range $0.001 \le K \le 0.016$ while the mean separation between particles is fixed at $L=32$ in Fig.~\ref{f:w2deK}(a) and $L=16$ in Fig.~\ref{f:w2deK}(b).

As can be observed, the three regimes are still present. The saturation value of the roughness $W^2_2$ decreases with increasing $K$. In addition, for a fixed $K$, $W^2_2$ grows with the mean separation $L$. In fact, this behavior can be rationalized by considering that when increasing the intensity $K$, the \textit{effective} mean separation between particles decreases since the harmonic interaction generates an effective collision before two consecutive particles actually touch each other. Therefore, this effective decrease of $L$ makes the saturation value of the roughness to decrease.

\begin{figure}[!tbp]
\includegraphics[width=8.5cm,clip=true]{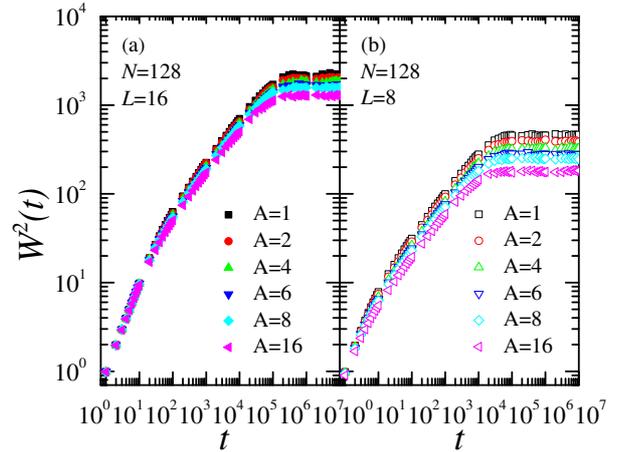}
\caption{\label{f:w2deA}(Color online) Behavior of the roughness for the case of an additional interaction term proportional to $1/r^2$ and for $N=128$. The mean separation is fixed at $L=16$ in panel (a) and $L=8$ in panel (b), with the different $A$ values indicated in the keys.}
\end{figure}

Figure~\ref{f:w2deA} shows the behavior of the roughness for an interaction term proportional to $1/r^2$. The number of particles is $N=128$ and the mean separation is fixed at $L=16$ in Fig.~\ref{f:w2deA}(a) and $L=8$ in Fig.~\ref{f:w2deA}(b). The intensity of the interaction takes values in the range $1 \le A \le 16$. The general trend is the same as in the harmonic case. For larger intensities, the saturation values decrease. This can also be rationalized in terms of a decreasing effective separation between particles. The fact that the separation between particles is effectively changing with the additional interaction terms affects all the crossover values, as we shal discuss in the next subsection.

\subsection{Crossover points and scaling functions}

We analyze here how the crossover points $y_1=(t_1,W^2_1)$ and $y_2=(t_2,W^2_2)$ depend on the interaction intensities $K$ and $A$. For clarity, let us call $y_1^0$ and $y_2^0$ the values these parameters take in the pure entropic case, given by Eqs.~\eqref{eq:y0}.

For an harmonic interaction between particles it is clear that when $K$ is very large particles will oscillate around their initial positions. Then, since two particles would hardly be in neighboring sites the hard core repulsion can be neglected in the large $K$ limit. The system will then be very well described by the DEWE with the elastic constant given directly by $\nu=K$. The crossover parameters $y_1$ and $y_2$ will take the values given by Eqs.~\eqref{eq:ynu1} and \eqref{eq:ynu2} with $\nu=K$ and $D=\Gamma$.

Therefore, the crossover parameters change from the $K=0$ value given by Eqs.~\eqref{eq:y0} to the large $K$ values given by Eqs.~\eqref{eq:ynu1} and \eqref{eq:ynu2} with $\nu=K$. One can expect the crossover between these values to be given in terms of the quotient $K/\nu_{\text{SFD}}$, since this is the effective variable which is changing. We then expect that the scaled values $y_1/y_1^0$ be only a function of $KL^2/D$; and the same for $y_2/y_2^0$. We can thus write that
\begin{equation}
\label{eq:scalyK}
 \widetilde y_m = \frac{y_m}{y_m^0} = f\left( \frac{KL^2}{D} \right),
\end{equation}
with $m=1,2$, and the scaling function behaving as
\begin{equation}
\label{eq:fK}
 f(x) = \left\{
 \begin{array}{lll}
  1 &\text{for}&x \ll 1, \\
  x^{-1}&\text{for}& x \gg 1.
 \end{array}
\right.
\end{equation}
The $x^{-1}$ behavior at large $K$ results directly from Eqs.~\eqref{eq:ynu1} and \eqref{eq:ynu2}. Furthermore, since the dependence of all the crossover parameters with $\nu$ is the same, we expect the same scaling function $f(x)$ for all of them.

\begin{figure}[!tbp]
\includegraphics[width=8.5cm,clip=true]{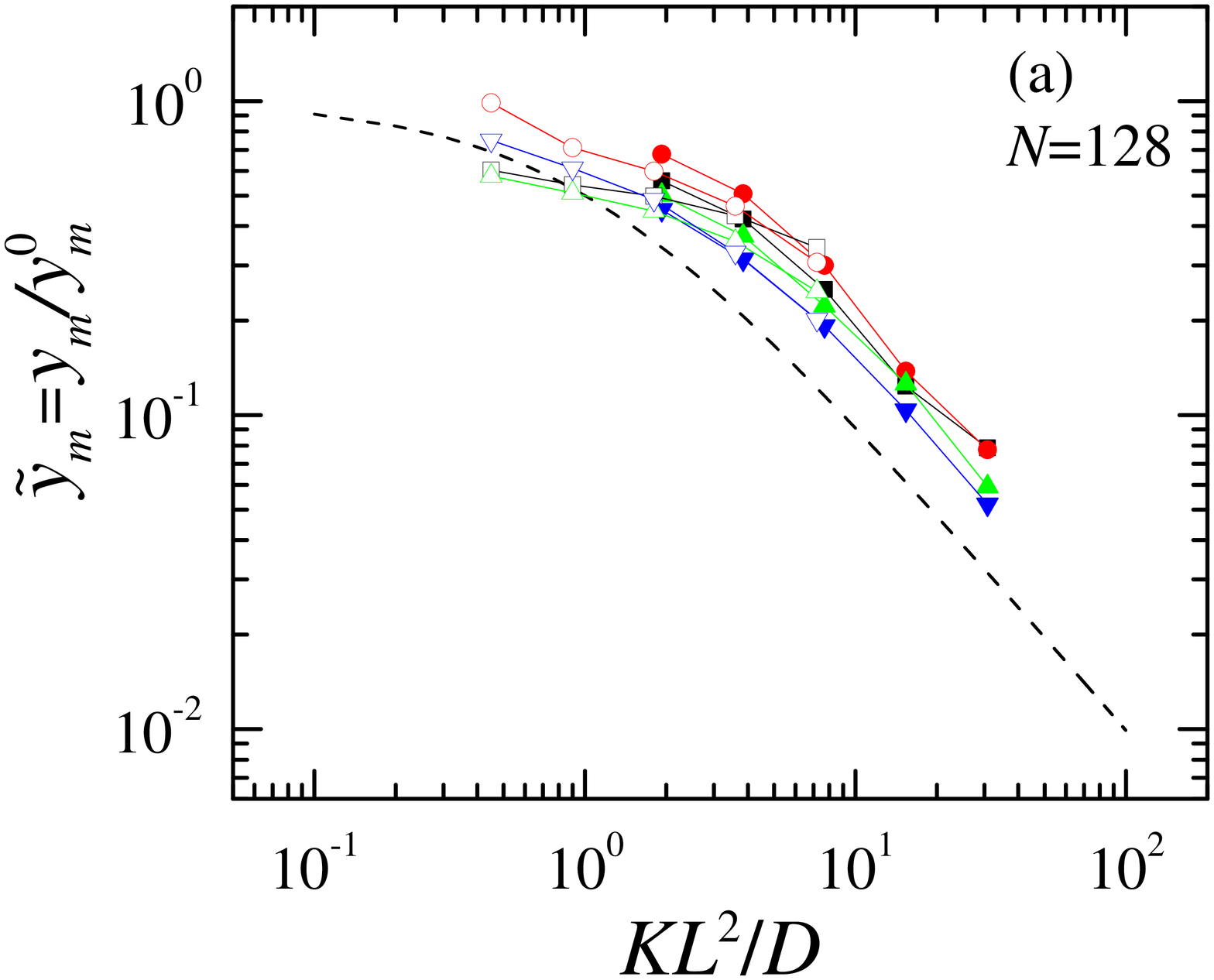}
\includegraphics[width=8.5cm,clip=true]{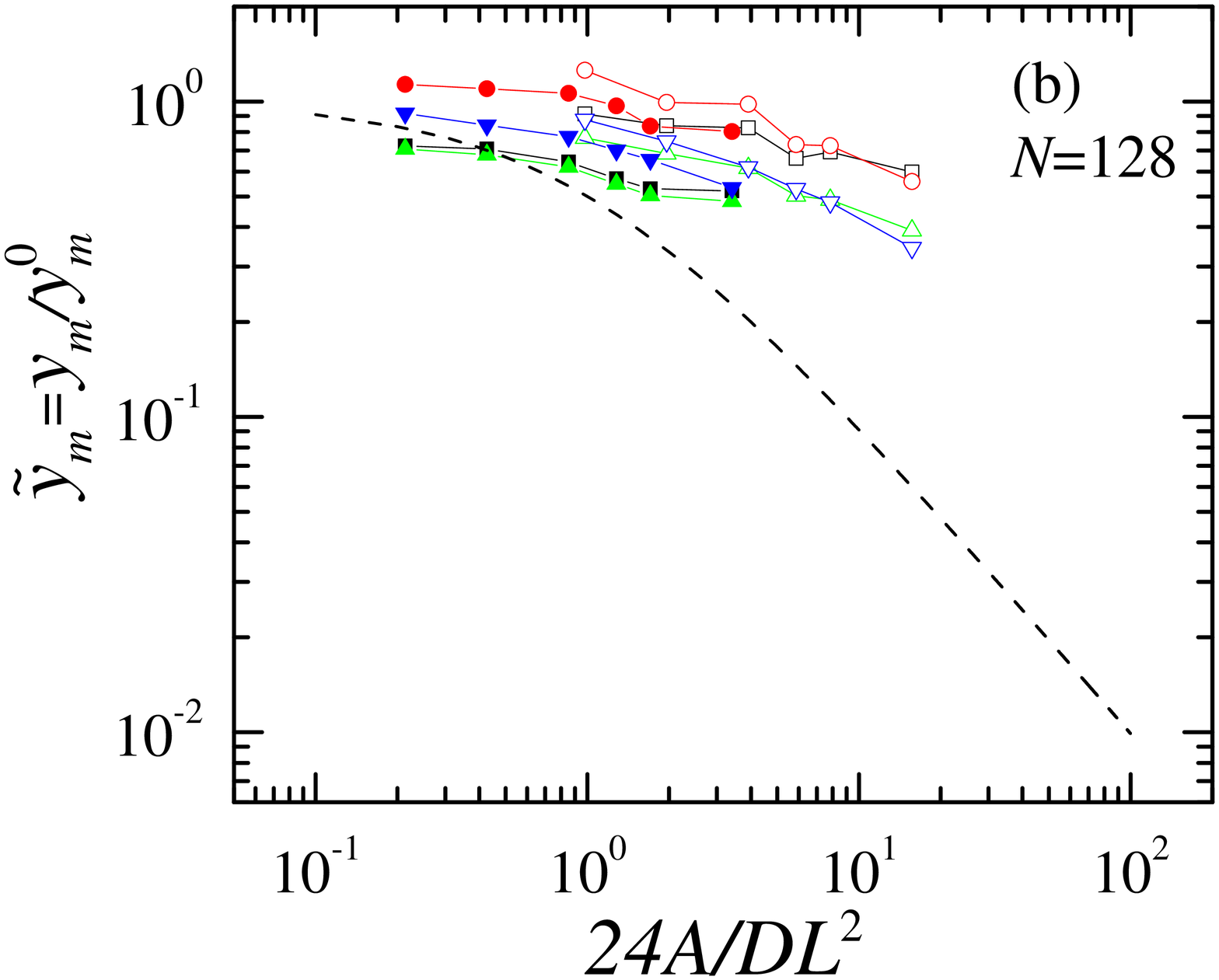}
\caption{\label{f:y1y2scal}(Color online) Scaling of the crossover points for the cases with additional interaction terms. (a) The harmonic case, see Eq.~\eqref{eq:scalyK}, with $L=32$ (closed symbols) and $L=16$ (open symbols). (b) The $1/r^2$ case, see Eq.~\eqref{eq:scalyA}, with $L=16$ (closed symbols) and $L=8$ (open symbols). In both panels the crossover parameters are: $t_1$ (squares), $t_2$ (circles), $W^2_1$ (triangles up), and $W^2_2$ (triangles down).}
\end{figure}

From the data in Fig.~\ref{f:w2deK} we have estimated the two crossover points for all curves, which are shown in Fig.~\ref{f:y1y2scal}(a) according to the proposed scaling form, Eq.~\eqref{eq:scalyK}. In this figure, close (open) symbols correspond to $L=32$ ($L=16$). The crossover parameters are: $t_1$ (squares), $t_2$ (circles), $W^2_1$ (triangles up), and $W^2_2$ (triangles down). As a guide to the eye we have also plotted in the figure, using a dashed line, a simple function that behaves as the scaling function Eq.\eqref{eq:fK}. This function is given by $f(x) = (1 + x)^{-1}$. Although the error bar of the estimated values of the crossover points is of the order of the scatter of all the points in the figure, one can fairly observe that the data follow the general trends of the proposed scaling form.

\begin{figure}[!tbp]
\includegraphics[width=8.5cm,clip=true]{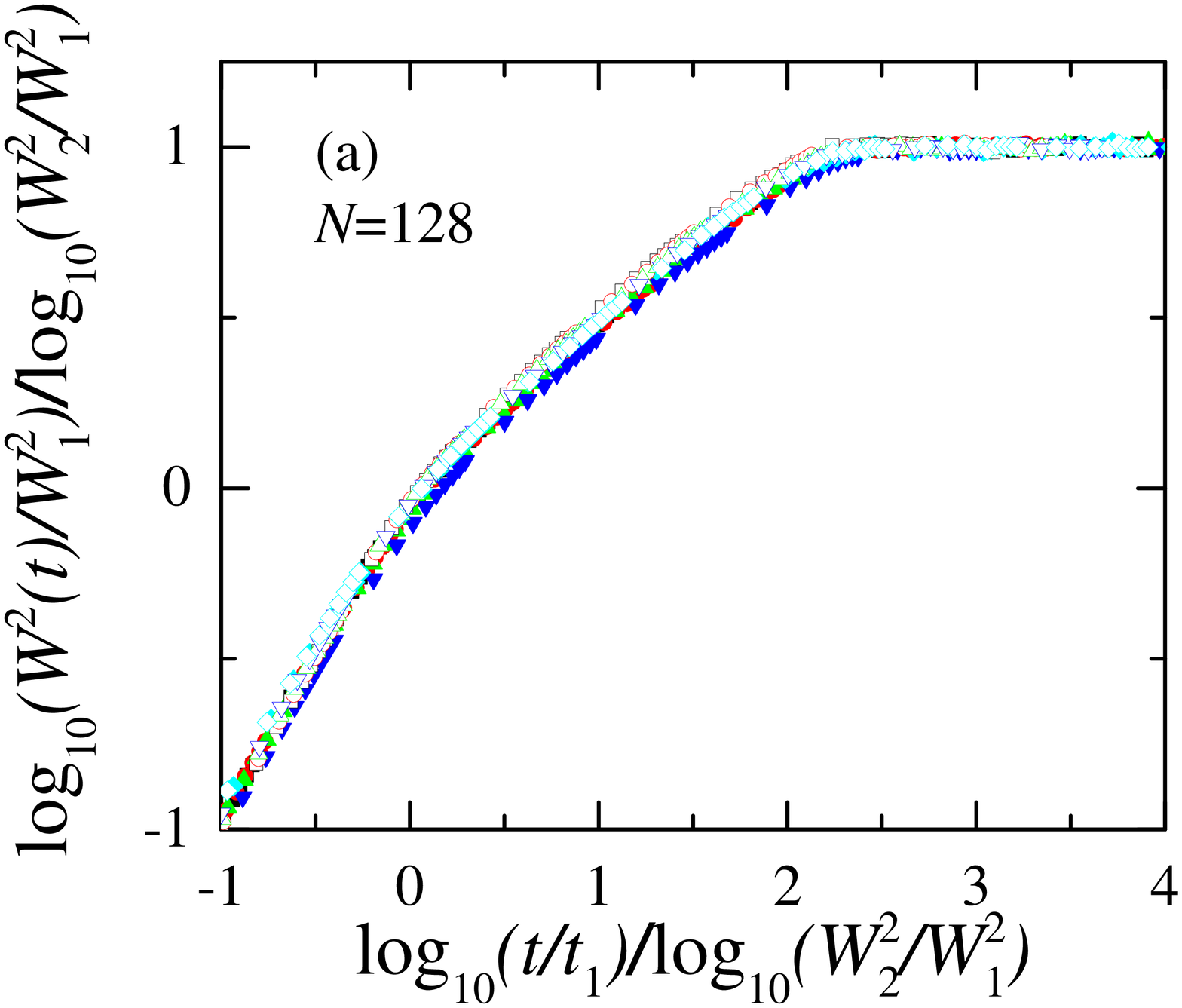}
\includegraphics[width=8.5cm,clip=true]{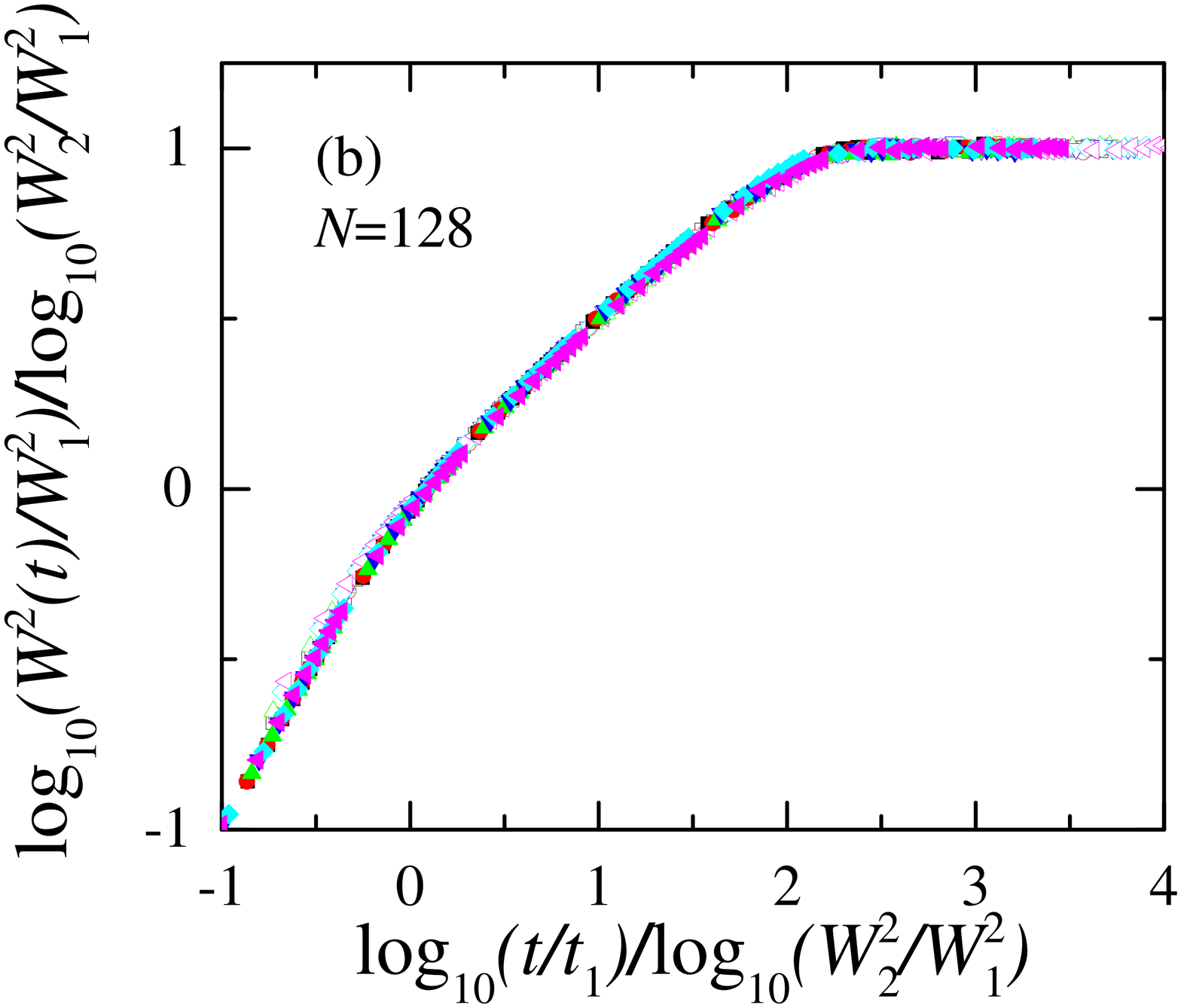}
\caption{\label{f:w2scalKA}(Color online) Roughness scaling for the (a) harmonic and (b) $1/r^2$ interaction terms.}
\end{figure}

In the case of the $1/r^2$ potential, it is convenient to consider the large $A$ limit performing a small $u$ expansion. To this end, let us consider a simplified situation where one particle can diffuse between its two neighboring particles which are at fixed positions, $+L$ and $-L$ respectively. Then, if $A$ is large, the particle will oscillate around $u=0$ and the energy can be approximated by
\begin{equation}
 \delta U \approx \frac{12 A}{L^4} \delta u^2 = \frac{\nu_A}{2} \delta u^2.
\end{equation}
Therefore, in an harmonic approximation, one can consider the effective elastic constant to be $\nu_A = 24 A/L^4$.

Since in the large $A$ limit the system can be considered within an harmonic approximation, the arguments we followed for the harmonic potential can be straightforwardly used here. In this case, the parameters $y_m$ change from the $y_m^0$ value to the values given by Eqs.~\eqref{eq:ynu1} and \eqref{eq:ynu2} but this time with $\nu_{\text{SFD}}=\nu_A$. It follows that
\begin{equation}
\label{eq:scalyA}
 \widetilde y_m = \frac{y_m}{y_m^0} = g\left( \frac{24 A}{D L^2} \right),
\end{equation}
with the scaling function behaving as
\begin{equation}
 g(x) = \left\{
 \begin{array}{lll}
  1 &\text{for}&x \ll 1, \\
  x^{-1}&\text{for}& x \gg 1.
 \end{array}
\right.
\end{equation}
It is also tempting to see if the two functions $f$ and $g$ are approximately the same, although this is not obvious \textit{a priori}.

From the data in Fig.~\ref{f:w2deA} we have estimated the values of the crossover points, as shown in the scaled form Eq.~\eqref{eq:scalyA} in Fig.~\ref{f:y1y2scal}(b). Closed (open) symbols correspond to $L=16$ ($L=8$) and the crossover parameters are plotted with the same symbols as in Fig.~\ref{f:y1y2scal}(a). The dashed line corresponds to $g(x) = (1+x)^{-1}$ and is plotted as a guide to the eye. As compared with the harmonic case, these data points seem to be closer to the saturation of $g(x)$ at very small values. In order to properly reach the large $A$ regime one should go to larger system sizes. Besides, it should be mentioned here that although we expect the large $A$ limit to be described with the scaling form Eq.~\eqref{eq:scalyA}, the small $A$ limit will not necessarily be described with this scaling form since the small $\delta u$ expansion is not valid in this limit.

Finally, since we have already estimated the values of the crossover points, we can also present the roughness scaling covering the three characteristic regimes by using the form given by Eq.~\eqref{eq:scal-chou}. Figure~\ref{f:w2scalKA} shows the roughness scaling for (a) the harmonic case and (b) the $1/r^2$ case. As can be observed, the data collapse in terms of the estimated crossover points is very good.

\section{Universal roughness distribution}

\begin{figure}[!tbp]
\includegraphics[width=8.5cm,clip=true]{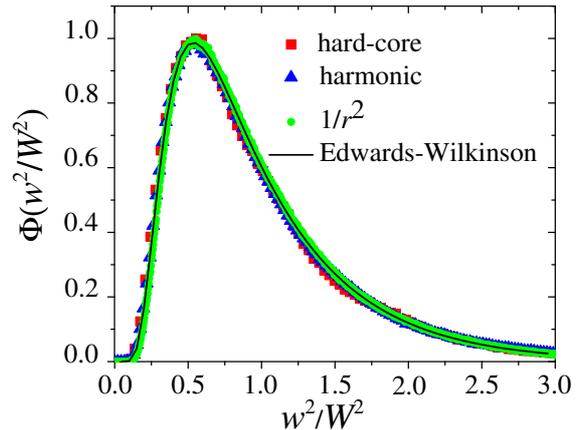}
\caption{\label{f:scalPhi}(Color online) Universal scaling of the roughness distribution for the pure hard-core case ($N=64$ and $L=8$), the harmonic potential ($N=64$, $L=32$ and $K=0.008$), and the $1/r^2$ potential ($N=64$, $L=8$ and $A=6$). The continuous line corresponds to the analytical solution of the Edwards-Wilkinson equation, from Ref.~\cite{foltin1994}.}
\end{figure}

As a final characterization of the roughness behavior for the SFD problem, we shall discuss here the form of the roughness distribution. As has been already shown, the roughness distribution can be used as a test of the surface growth universality class~\cite{racz1994,plischke1994,marinari2002,racz2003}. For instance, let $w^2$ be a given instantaneous value of the roughness at saturation, i.e. for $t \gg t_2$. The roughness distribution $P(w^2)$ can then be written in the scaling form~\cite{racz1994}
\begin{equation}
\label{eq:scalPhi}
 P(w^2) = \frac{1}{W^2} \Phi \left( \frac{w^2}{W^2} \right),
\end{equation}
where $W^2$ is the average value of the global roughness at saturation and thus corresponds to the mean value of $P(w^2)$. The function $\Phi(x)$ is a given scaling function characteristic of the universality class. The form of $\Phi(x)$ for the Edwards-Wilkinson universality class has been analytically calculated in Ref.~\cite{foltin1994} in the saturation regime and in Ref.~\cite{antal1996} in the correlated growing regime. In Fig.~\ref{f:scalPhi} we show the scaled roughness distribution for three data sets corresponding to the purely hard-core, harmonic, and $1/r^2$ repulsion cases. The analytical form of $\Phi(x)$ for the Edwards-Wilkinson case is also shown. As can be observed all three cases are very well collapsed into the same scaling function according to Eq.~\eqref{eq:scalPhi}.

\section{Concluding remarks}

Summarizing, we have shown that the DEWE with an effective elasticity $\nu_{\text{SFD}}=D/(L-1)^2$ describes the SFD process with hard-core repulsion. We have given a description of the process in terms of the roughness, which is a global measure of particle fluctuations. With the characteristic values of the roughness at the two crossover times $t_1$ and $t_2$, we were able to scale the roughness in terms of a scaling function showing the three regimes. Although the phenomenological scaling form works very well, we note that it is not clear how it is present in the exact solution for the DEWE.

We have as well explored the effects of different additional repulsive interaction potentials between the particles. For large intensity values, the dynamics of both the harmonic and $1/r^2$ potentials cases can be described in terms of the DEWE with an effective elasticity $K$ and $\nu_A$ respectively. We have shown how the values of the crossover points scale with the effective elasticity and we have properly scaled the global roughness.

Finally, we have shown that for all cases the roughness distribution in the saturation regime can be recast into the universal form of the Edwards-Wilkinson equation. This indicates that all the relevant parameters of the SFD process are taken into account by the mean value of the roughness.

We have offered here a simple description of the SFD process using the roughness of the associated effective DEWE, providing an alternative tool to describe the SFD problem, which can be complementary to other analytical approaches like the ones presented in Refs.~\cite{lizana2008,barkai2009}. Furthermore, the use of the effective DEWE opens up the possibility of exploring the problem from a new perspective. For instance, related equations also used in surface growth phenomena, like the Kardar-Parisi-Zhang or the Mullins-Herring equations~\cite{barabasi-stanley}, could be potentially used to describe processes when the non-passing rule is slightly relaxed. In this cases an effective mass transport could be considered using these equations. It can be also important to consider the effective Edwards-Wilkinson equation with quenched disorder to describe SFD processes in a disordered substrate as the one considered in Ref.~\cite{ben-naim2009}. Thus, we expect that the potential relation between SFD processes and surface growth equations can be further exploited.

Regarding the experimental relevance of our results, one should note that the interplay between SFD and the DEWE could also be exploited. In the field of vecinal surfaces it is well known that transverse fluctuations can be described with and Edwards-Wilkinson equation, but it is also accepted that the statistical distribution of the separation between surface steps is typically given by a Wigner surmise, commonly observed in SFD problems~\cite{giesen2001}. Thus, the relation presented here between SFD processes and the Edwards-Wilkinson equation could give new insights in this and related physical systems.

\acknowledgments

PMC acknowledges the INTER-U program (Secretar\'{\i}a de Pol\'{\i}ticas Universitarias-Miniterio de Educaci\'on, (Argentina) for financial support. This work was also supported in part by CONICET (Argentina) under project numbers PIP 5596 and PIP 112-200801-01332, Universidad Nacional de San Luis (Argentina) under project 32200, and the National Agency of Scientific and Technological Promotion (Argentina) under project 33328 PICT 2005.

\bibliography{steps}

\begin{thebibliography}{27}
\expandafter\ifx\csname natexlab\endcsname\relax\def\natexlab#1{#1}\fi
\expandafter\ifx\csname bibnamefont\endcsname\relax
  \def\bibnamefont#1{#1}\fi
\expandafter\ifx\csname bibfnamefont\endcsname\relax
  \def\bibfnamefont#1{#1}\fi
\expandafter\ifx\csname citenamefont\endcsname\relax
  \def\citenamefont#1{#1}\fi
\expandafter\ifx\csname url\endcsname\relax
  \def\url#1{\texttt{#1}}\fi
\expandafter\ifx\csname urlprefix\endcsname\relax\def\urlprefix{URL }\fi
\providecommand{\bibinfo}[2]{#2}
\providecommand{\eprint}[2][]{\url{#2}}

\bibitem[{\citenamefont{Harris}(1965)}]{harris1965}
\bibinfo{author}{\bibfnamefont{T.~E.} \bibnamefont{Harris}},
  \bibinfo{journal}{J. Appl. Prob.} \textbf{\bibinfo{volume}{2}},
  \bibinfo{pages}{323} (\bibinfo{year}{1965}).

\bibitem[{\citenamefont{Fedders}(1978)}]{fedders1978}
\bibinfo{author}{\bibfnamefont{P.~A.} \bibnamefont{Fedders}},
  \bibinfo{journal}{Phys. Rev. B} \textbf{\bibinfo{volume}{17}},
  \bibinfo{pages}{40} (\bibinfo{year}{1978}).

\bibitem[{\citenamefont{Alexander and Pincus}(1978)}]{alexander1978}
\bibinfo{author}{\bibfnamefont{S.}~\bibnamefont{Alexander}} \bibnamefont{and}
  \bibinfo{author}{\bibfnamefont{P.}~\bibnamefont{Pincus}},
  \bibinfo{journal}{Phys. Rev. B} \textbf{\bibinfo{volume}{18}},
  \bibinfo{pages}{2011} (\bibinfo{year}{1978}).

\bibitem[{\citenamefont{van Beijeren et~al.}(1983)\citenamefont{van Beijeren,
  Kehr, and Kutner}}]{beijeren1983}
\bibinfo{author}{\bibfnamefont{H.}~\bibnamefont{van Beijeren}},
  \bibinfo{author}{\bibfnamefont{K.~W.} \bibnamefont{Kehr}}, \bibnamefont{and}
  \bibinfo{author}{\bibfnamefont{R.}~\bibnamefont{Kutner}},
  \bibinfo{journal}{Phys. Rev. B} \textbf{\bibinfo{volume}{28}},
  \bibinfo{pages}{5711} (\bibinfo{year}{1983}).

\bibitem[{\citenamefont{Lin et~al.}(2005)\citenamefont{Lin, Meron, Cui, Rice,
  and Diamant}}]{lin2005}
\bibinfo{author}{\bibfnamefont{B.}~\bibnamefont{Lin}},
  \bibinfo{author}{\bibfnamefont{M.}~\bibnamefont{Meron}},
  \bibinfo{author}{\bibfnamefont{B.}~\bibnamefont{Cui}},
  \bibinfo{author}{\bibfnamefont{S.~A.} \bibnamefont{Rice}}, \bibnamefont{and}
  \bibinfo{author}{\bibfnamefont{H.}~\bibnamefont{Diamant}},
  \bibinfo{journal}{Phys. Rev. Lett.} \textbf{\bibinfo{volume}{94}},
  \bibinfo{pages}{216001} (\bibinfo{year}{2005}).

\bibitem[{\citenamefont{Lizana and Ambj\"ornsson}(2008)}]{lizana2008}
\bibinfo{author}{\bibfnamefont{L.}~\bibnamefont{Lizana}} \bibnamefont{and}
  \bibinfo{author}{\bibfnamefont{T.}~\bibnamefont{Ambj\"ornsson}},
  \bibinfo{journal}{Phys. Rev. Lett.} \textbf{\bibinfo{volume}{100}},
  \bibinfo{pages}{200601} (\bibinfo{year}{2008}).

\bibitem[{\citenamefont{Lizana and Ambj\"ornsson}(2009)}]{lizana2009}
\bibinfo{author}{\bibfnamefont{L.}~\bibnamefont{Lizana}} \bibnamefont{and}
  \bibinfo{author}{\bibfnamefont{T.}~\bibnamefont{Ambj\"ornsson}},
  \bibinfo{journal}{Phys. Rev. E} \textbf{\bibinfo{volume}{80}},
  \bibinfo{pages}{051103} (\bibinfo{year}{2009}).

\bibitem[{\citenamefont{Taloni and Lomholt}(2008)}]{taloni2008}
\bibinfo{author}{\bibfnamefont{A.}~\bibnamefont{Taloni}} \bibnamefont{and}
  \bibinfo{author}{\bibfnamefont{M.~A.} \bibnamefont{Lomholt}},
  \bibinfo{journal}{Phys. Rev. E} \textbf{\bibinfo{volume}{78}},
  \bibinfo{pages}{051116} (\bibinfo{year}{2008}).

\bibitem[{\citenamefont{Barkai and Silbey}(2009)}]{barkai2009}
\bibinfo{author}{\bibfnamefont{E.}~\bibnamefont{Barkai}} \bibnamefont{and}
  \bibinfo{author}{\bibfnamefont{R.}~\bibnamefont{Silbey}},
  \bibinfo{journal}{Phys. Rev. Lett.} \textbf{\bibinfo{volume}{102}},
  \bibinfo{pages}{050602} (\bibinfo{year}{2009}).

\bibitem[{\citenamefont{Wei et~al.}(1999)\citenamefont{Wei, Bechinger, and
  Leiderer}}]{wei1999}
\bibinfo{author}{\bibfnamefont{Q.-H.} \bibnamefont{Wei}},
  \bibinfo{author}{\bibfnamefont{C.}~\bibnamefont{Bechinger}},
  \bibnamefont{and} \bibinfo{author}{\bibfnamefont{P.}~\bibnamefont{Leiderer}},
  \bibinfo{journal}{Prog. Colloid. Polym. Sci.} \textbf{\bibinfo{volume}{122}},
  \bibinfo{pages}{227} (\bibinfo{year}{1999}).

\bibitem[{\citenamefont{Wei et~al.}(2000)\citenamefont{Wei, Bechinger, and
  Leiderer}}]{wei2000}
\bibinfo{author}{\bibfnamefont{Q.-H.} \bibnamefont{Wei}},
  \bibinfo{author}{\bibfnamefont{C.}~\bibnamefont{Bechinger}},
  \bibnamefont{and} \bibinfo{author}{\bibfnamefont{P.}~\bibnamefont{Leiderer}},
  \bibinfo{journal}{Science} \textbf{\bibinfo{volume}{287}},
  \bibinfo{pages}{625} (\bibinfo{year}{2000}).

\bibitem[{\citenamefont{Coupier et~al.}(2006)\citenamefont{Coupier, {Saint
  Jean}, and Guthmann}}]{coupier2006}
\bibinfo{author}{\bibfnamefont{G.}~\bibnamefont{Coupier}},
  \bibinfo{author}{\bibfnamefont{M.}~\bibnamefont{{Saint Jean}}},
  \bibnamefont{and} \bibinfo{author}{\bibfnamefont{C.}~\bibnamefont{Guthmann}},
  \bibinfo{journal}{Phys. Rev. E} \textbf{\bibinfo{volume}{73}},
  \bibinfo{pages}{031112} (\bibinfo{year}{2006}).

\bibitem[{\citenamefont{Amitai et~al.}(2010)\citenamefont{Amitai, Kantor, and
  Kardar}}]{amitai2010}
\bibinfo{author}{\bibfnamefont{A.}~\bibnamefont{Amitai}},
  \bibinfo{author}{\bibfnamefont{Y.}~\bibnamefont{Kantor}}, \bibnamefont{and}
  \bibinfo{author}{\bibfnamefont{M.}~\bibnamefont{Kardar}},
  \bibinfo{journal}{Phys. Rev. E} \textbf{\bibinfo{volume}{81}},
  \bibinfo{pages}{011107} (\bibinfo{year}{2010}).

\bibitem[{\citenamefont{Doi and Edwards}(1986)}]{doi1986}
\bibinfo{author}{\bibfnamefont{M.}~\bibnamefont{Doi}} \bibnamefont{and}
  \bibinfo{author}{\bibfnamefont{S.~F.} \bibnamefont{Edwards}},
  \emph{\bibinfo{title}{The theory of polymer dynamics}}
  (\bibinfo{publisher}{Clarendon Press}, \bibinfo{address}{Oxford},
  \bibinfo{year}{1986}).

\bibitem[{\citenamefont{Giesen}(2001)}]{giesen2001}
\bibinfo{author}{\bibfnamefont{M.}~\bibnamefont{Giesen}},
  \bibinfo{journal}{Prog. Surf. Science} \textbf{\bibinfo{volume}{68}},
  \bibinfo{pages}{1} (\bibinfo{year}{2001}).

\bibitem[{\citenamefont{Lim and Teo}(2009)}]{lim2009}
\bibinfo{author}{\bibfnamefont{S.~C.} \bibnamefont{Lim}} \bibnamefont{and}
  \bibinfo{author}{\bibfnamefont{L.~P.} \bibnamefont{Teo}},
  \bibinfo{journal}{J. Stat. Mech.: Theor. Exp.} p. \bibinfo{pages}{P08015}
  (\bibinfo{year}{2009}).

\bibitem[{\citenamefont{Edwards and Wilkinson}(1982)}]{edwards1982}
\bibinfo{author}{\bibfnamefont{S.~F.} \bibnamefont{Edwards}} \bibnamefont{and}
  \bibinfo{author}{\bibfnamefont{D.~R.} \bibnamefont{Wilkinson}},
  \bibinfo{journal}{Proc. R. Soc. A} \textbf{\bibinfo{volume}{381}},
  \bibinfo{pages}{17} (\bibinfo{year}{1982}).

\bibitem[{\citenamefont{Bar\'abasi and Stanley}(1995)}]{barabasi-stanley}
\bibinfo{author}{\bibfnamefont{A.-L.} \bibnamefont{Bar\'abasi}}
  \bibnamefont{and} \bibinfo{author}{\bibfnamefont{H.~E.}
  \bibnamefont{Stanley}}, \emph{\bibinfo{title}{Fractal concepts in surface
  growth}} (\bibinfo{publisher}{Cambridge university Press},
  \bibinfo{address}{Cambridge, England}, \bibinfo{year}{1995}).

\bibitem[{\citenamefont{Family and Vicsek}(1985)}]{family1985}
\bibinfo{author}{\bibfnamefont{F.}~\bibnamefont{Family}} \bibnamefont{and}
  \bibinfo{author}{\bibfnamefont{T.}~\bibnamefont{Vicsek}},
  \bibinfo{journal}{J. Phys. A} \textbf{\bibinfo{volume}{18}},
  \bibinfo{pages}{L75} (\bibinfo{year}{1985}).

\bibitem[{\citenamefont{Chou and Pleimling}(2009)}]{chou2009}
\bibinfo{author}{\bibfnamefont{Y.-L.} \bibnamefont{Chou}} \bibnamefont{and}
  \bibinfo{author}{\bibfnamefont{M.}~\bibnamefont{Pleimling}},
  \bibinfo{journal}{Phys. Rev. E} \textbf{\bibinfo{volume}{79}},
  \bibinfo{pages}{051605} (\bibinfo{year}{2009}).

\bibitem[{\citenamefont{Foltin et~al.}(1994)\citenamefont{Foltin, Oerding,
  R\'acz, Workman, and Zia}}]{foltin1994}
\bibinfo{author}{\bibfnamefont{G.}~\bibnamefont{Foltin}},
  \bibinfo{author}{\bibfnamefont{K.}~\bibnamefont{Oerding}},
  \bibinfo{author}{\bibfnamefont{Z.}~\bibnamefont{R\'acz}},
  \bibinfo{author}{\bibfnamefont{R.~L.} \bibnamefont{Workman}},
  \bibnamefont{and} \bibinfo{author}{\bibfnamefont{R.~K.~P.}
  \bibnamefont{Zia}}, \bibinfo{journal}{Phys. Rev. E}
  \textbf{\bibinfo{volume}{50}}, \bibinfo{pages}{R639} (\bibinfo{year}{1994}).

\bibitem[{\citenamefont{R\'acz and Plischke}(1994)}]{racz1994}
\bibinfo{author}{\bibfnamefont{Z.}~\bibnamefont{R\'acz}} \bibnamefont{and}
  \bibinfo{author}{\bibfnamefont{M.}~\bibnamefont{Plischke}},
  \bibinfo{journal}{Phys. Rev. E} \textbf{\bibinfo{volume}{50}},
  \bibinfo{pages}{3530} (\bibinfo{year}{1994}).

\bibitem[{\citenamefont{Plischke et~al.}(1994)\citenamefont{Plischke, R\'acz,
  and Zia}}]{plischke1994}
\bibinfo{author}{\bibfnamefont{M.}~\bibnamefont{Plischke}},
  \bibinfo{author}{\bibfnamefont{Z.}~\bibnamefont{R\'acz}}, \bibnamefont{and}
  \bibinfo{author}{\bibfnamefont{R.~K.~P.} \bibnamefont{Zia}},
  \bibinfo{journal}{Phys. Rev. E} \textbf{\bibinfo{volume}{50}},
  \bibinfo{pages}{3589} (\bibinfo{year}{1994}).

\bibitem[{\citenamefont{Marinari et~al.}(2002)\citenamefont{Marinari, Pagnani,
  Parisi, and R\'acz}}]{marinari2002}
\bibinfo{author}{\bibfnamefont{E.}~\bibnamefont{Marinari}},
  \bibinfo{author}{\bibfnamefont{A.}~\bibnamefont{Pagnani}},
  \bibinfo{author}{\bibfnamefont{G.}~\bibnamefont{Parisi}}, \bibnamefont{and}
  \bibinfo{author}{\bibfnamefont{Z.}~\bibnamefont{R\'acz}},
  \bibinfo{journal}{Phys. Rev. E} \textbf{\bibinfo{volume}{65}},
  \bibinfo{pages}{026136} (\bibinfo{year}{2002}).

\bibitem[{\citenamefont{R\'acz}(2003)}]{racz2003}
\bibinfo{author}{\bibfnamefont{Z.}~\bibnamefont{R\'acz}},
  \bibinfo{journal}{SPIE Proc.} \textbf{\bibinfo{volume}{5112}},
  \bibinfo{pages}{248} (\bibinfo{year}{2003}).

\bibitem[{\citenamefont{Antal and R\'acz}(1996)}]{antal1996}
\bibinfo{author}{\bibfnamefont{T.}~\bibnamefont{Antal}} \bibnamefont{and}
  \bibinfo{author}{\bibfnamefont{Z.}~\bibnamefont{R\'acz}},
  \bibinfo{journal}{Phys. Rev. E} \textbf{\bibinfo{volume}{54}},
  \bibinfo{pages}{2256} (\bibinfo{year}{1996}).

\bibitem[{\citenamefont{Ben-Naim and Krapivsky}(2009)}]{ben-naim2009}
\bibinfo{author}{\bibfnamefont{E.}~\bibnamefont{Ben-Naim}} \bibnamefont{and}
  \bibinfo{author}{\bibfnamefont{P.~L.} \bibnamefont{Krapivsky}},
  \bibinfo{journal}{Phys. Rev. Lett.} \textbf{\bibinfo{volume}{102}},
  \bibinfo{pages}{190602} (\bibinfo{year}{2009}).

\end{thebibliography}

\end{document}